\documentclass[sigplan,nonacm]{acmart}
\AtBeginDocument{%
  }

\newcommand{\cherid}{\mbox{CHERI-\textbf{\textit{D}}}}
\newcommand{\reinc}{\mbox{\textbf{\textit{Reinc}}}}

\usepackage{multirow}
\usepackage{algorithm}
\usepackage{algpseudocode}
\usepackage{balance}
\usepackage{booktabs}
\usepackage{subcaption}
\usepackage{svg}

\begin{document}

\title[CHERI-D Reincarnate]{
CHERI-\textbf{\textit{D}} Reincarnate:
efficient multicore CHERI temporal memory safety through allocation reincarnation (draft version)
}
\author{Yuecheng Wang}
\orcid{0009-0009-8075-7006}
\affiliation{%
  \institution{University of Cambridge}
  \country{United Kingdom}
}
\email{yuecheng.wang@cl.cam.ac.uk}

\author{Jonathan Woodruff}
\affiliation{%
  \institution{University of Cambridge}
  \country{United Kingdom}
}
\email{jonathan.woodruff@cl.cam.ac.uk}

\author{Simon W. Moore}
\affiliation{%
  \institution{University of Cambridge}
  \country{United Kingdom}
}
\orcid{0000-0002-2806-495X}
\email{simon.moore@cl.cam.ac.uk}

\renewcommand{\shortauthors}{Wang et al.}

\begin{abstract}
We propose CHERI-\textbf{\textit{D}} Reincarnate (\reinc{}), an architectural extension to CHERI for scalable and efficient temporal memory safety. 
Prior work \cherid{} has a finite-width generation ID stored at a fixed location, requiring an object to be quarantined when
its ID is exhausted.
\reinc{} further provides use-after-free mitigation while permitting immediate freed memory reuse for objects through \emph{allocation reincarnation}: rather than quarantining an allocation slot upon ID exhaustion, \reinc{} dynamically assigns a new ID to that slot when its current ID is exhausted. Exhausted IDs are quarantined and later reclaimed, while the underlying memory remains available for immediate reuse. By quarantining IDs rather than memory, \reinc{} enables continuous reuse of memory in the common case,
substantially reducing both memory-sweep frequency and quarantine memory overhead.

\reinc{} further introduces coherent ID caching while retaining a fully decentralized ID organization. Temporal metadata remains colocated with the memory it protects, preserving locality while avoiding centralized metadata structures. 
To support multicore execution, \reinc{} connects physical coherence events to the virtually addressed ObjID buffer using lightweight reverse-map and filter-based mechanisms.

We implement \reinc{} as a hardware-software co-design spanning CHERI-Toooba (superscalar FPGA softcore), QEMU, LLVM/Clang and CheriBSD~\cite{ruggSuiteProcessorsExplore2024, qemu, llvm, cheribsd}. Across our evaluated workloads, \reinc{} substantially reduces memory-sweep frequency and memory quarantine while incurring low performance and hardware overhead.

\end{abstract}






\maketitle

\section{Introduction}
CHERI provides fine-grained spatial memory safety by replacing conventional pointers with hardware-protected capabilities carrying bounds, permissions, and provenance~\cite{cheriv9}. 
Temporal safety remains more challenging: capabilities can be freely copied and may remain reachable after the lifetime of the allocation they reference has ended.

The deployed CHERI temporal safety solution, Cornucopia Reloaded~\cite{Cornucopia_reloaded}, employs a software-based approach that delays reallocation until a sweep of the address space has revoked dangling heap capabilities.
Unfortunately, memory quarantining can significantly affect allocator optimization, memory footprint, and cache efficiency~\cite{wangPoisonCapEfficientHierarchical2026, wesleyfilardoCornucopiaTemporalSafety2020}.
When added to the cost of revocation sweeps, overall system performance and resource usage can be prohibitively impacted.

Recent work, including Picasso\cite{gulmezPICASSOScalingCHERI2026} and \cherid{} (pronounced, ``cherry-dee"), has demonstrated support for stronger temporal safety guarantees at much greater efficiency~\cite{wangCHERID2026}.
Picasso tracks object validity in a global bit-vector table indexed by a large object ID field in each capability,
and \cherid{} places an 8-bit generation ID in line with allocations that are smaller than one page.
Both provide temporal guarantees by checking metadata on memory access; capabilities pointing to freed objects are instantly revoked by clearing the validity bit (Picasso) or incrementing the object ID (\cherid{}).

\cherid{} has several advantages over Picasso; placing IDs inline with allocation data makes the approach more scalable, and counters use metadata memory more efficiently than bit vectors.
However, we see several opportunities for \cherid{} to be improved.
While \cherid{} allowed expressing a range of ID locations, it associated IDs statically with allocation slots, unnecessarily forcing memory to be quarantined upon ID exhaustion.
In addition, \cherid{} only supported objects of less than 4 KiB; while small objects are by far the most common, large objects disproportionately contribute to quarantine wastage and dominate many applications~\cite{wangCHERID2026}.
Finally, both Picasso and \cherid{} have been evaluated only under a single-core setup due to the complexity of synchronizing the virtually addressed metadata buffer across multiple cores. 

We propose \cherid{} Reincarnate (\reinc{}), an architectural extension to CHERI and a hardware-software co-design for supporting temporal safety for application-class multicore systems and long-running, allocation-intensive workloads. 
The key observation behind \reinc{} is that \emph{exhausting an identity need not exhaust the
memory it protects}. 
Instead, \reinc{} quarantines the exhausted ID and reincarnates the allocation slot under another available ID, allowing the underlying memory to remain immediately reusable. 
Quarantined IDs can subsequently be reclaimed after a memory sweep that invalidates outstanding capabilities carrying those IDs.
If reclamation completes before all available ID locations have been exhausted, the memory itself never needs to enter quarantine.

\reinc{} further extends ID-based temporal protection from 4 KiB to 1 GiB and addresses the multicore coherence problem introduced by private ObjID buffers. 
It keeps cached ObjIDs coherent across cores using lightweight mechanisms that reuse existing cache-coherence invalidation and L1-replacement events. 

Our evaluation shows that \reinc{} incurs average runtime overheads of 1.7\% for SPEC CPU2006 INT, 0.7\% for SQLite, and 0.3\% for PARSEC, while substantially reducing sweeping and memory quarantine overhead; for example, the SQLite benchmark requires no memory sweeps under \reinc{}, compared with 267 under Cornucopia Reloaded.

In this work, we make the following contributions:
\begin{itemize}
    \item We introduce \emph{allocation reincarnation} and \emph{ID quarantine and reclamation}, decoupling ID exhaustion from memory quarantine by reclaiming exhausted IDs independently of their underlying allocations.
    \item We extend ID-based temporal protection to objects up to 1 GiB and design two lightweight coherence mechanisms for synchronizing private ObjID buffers across cores using existing cache-coherence and L1-replacement events.
    \item We implement \reinc{} as a complete hardware-software co-design spanning CHERI-Toooba, QEMU, LLVM/Clang, CheriBSD, MRS, and jemalloc, demonstrating strict UAF mitigation with immediate memory reuse on a multicore CHERI system.
    \item We evaluate \reinc{} using SPEC CPU2006 INT, PARSEC, SQLite, NIST Juliet, and MSET, demonstrating strict UAF mitigation while substantially reducing memory-sweep frequency and memory quarantine.

\end{itemize}

\section{Background}
\subsection{CHERI}

CHERI extends conventional architectures with hardware-enforced \emph{capabilities}, which replace ordinary pointers with protected references carrying bounds, permissions, and provenance information.
Memory accesses through a capability are constrained to its authorized address range and permissions, providing fine-grained spatial memory safety.
CHERI follows a decentralized protection model: capabilities carry their access authority directly, avoiding centralized pointer metadata and allowing protection state to follow the existing memory hierarchy.
However, CHERI capabilities do not inherently encode object lifetime: a capability may remain valid after the object it references has been freed, motivating several hardware-software temporal-safety mechanisms.

\subsection{Cornucopia Reloaded}
Cornucopia Reloaded~\cite{Cornucopia_reloaded} enforces temporal safety for the heap using quarantine and revocation. Freed allocations are quarantined until a revocation sweep identifies and revokes dangling capabilities before the memory can be reused. 
This avoids adding metadata lookup to common-case memory accesses.
However, quarantine delays memory reuse and incurs performance and memory overheads for sometimes-frequent memory sweeps. It also does not provide strict use-after-free protection between deallocation and reuse.

\subsection{Picasso}
Picasso introduces a centralized table to record object validity, storing an object ID in each capability pointer and validating the corresponding bit of the table on each memory access\cite{gulmezPICASSOScalingCHERI2026}.
This mechanism provides immediate use-after-free protection and often has very low performance overhead.
However, centralized allocation of object IDs causes unexpected overheads; ID exhaustion can result in frequent sweeps; and table fragmentation can result in poor ID table buffering performance.

\subsection{\cherid{}}
\label{sec:background-cherid}

Following CHERI's decentralized protection model, \cherid{} distributes ID metadata inline with allocations, relying on CHERI bounds to preserve integrity~\cite{wangCHERID2026}.
Each capability records the location of its allocation's ObjID, which can often reside in unused fragmentation and incur no additional memory overhead. 
On each memory access, hardware compares the capability's ObjID with the corresponding memory ObjID and traps on a mismatch.

On deallocation, \cherid{} advances the memory ID before returning the object to the allocator, immediately invalidating capabilities from the previous generation. 
However, finite-width IDs eventually become exhausted after repeated reuse, at which point \cherid{} marks the ID as exhausted and falls back to memory quarantine.
An ID-aware memory sweep subsequently invalidates outstanding capabilities associated with exhausted IDs, allowing them to be reclaimed. This combination of decentralized ID metadata, immediate allocation slot reuse, and ID-aware invalidation sweeps forms the basis for \reinc{}.

\subsection{Temporal metadata buffer}
To reduce the latency of repeated ID accesses, Picasso introduces a color buffer~\cite{gulmezPICASSOScalingCHERI2026}, and \cherid{} introduces an object-ID (ObjID) buffer~\cite{wangCHERID2026} to cache recently accessed temporal metadata. 
In particular, \cherid{} relies on the ObjID buffer to cache IDs to allow most memory accesses to validate the object ID without an additional load for the metadata.
The ObjID buffer effectively operates as a virtually addressed cache, enabling ID lookup to proceed directly from the virtual address without waiting for address translation. 
This is important because ObjID validation lies on the latency-critical memory-access path; a virtually addressed buffer avoids both translation and memory access for ObjIds in the common case.

However, \cherid{}'s ObjID buffer and Picasso's color buffer operate independently of the processor's conventional cache-coherence hierarchy.
An ID update in memory, therefore, does not naturally invalidate or update a corresponding entry in the ObjID buffer. This problem is compounded by its virtually addressed organization: conventional cache-coherence protocols operate on physical addresses, whereas ObjID-buffer entries are identified using virtual addresses. 
A physical addresses from cache coherence messages, therefore, cannot be directly matched to a particular ObjID-buffer entry, while virtual aliases further prevent a simple one-to-one correspondence between virtual and physical addresses. 
This limitation becomes critical for multicore temporal safety. 
A remote core may continue to retain the previous ID in its private ObjID buffer after another core has updated the corresponding memory ID. 
A stale capability can therefore continue to match the obsolete locally cached ID and may not reliably trap. Picasso and the original \cherid{} design have consequently been evaluated only in single-core configurations with single-threaded workloads. Supporting multicore execution requires bridging physical cache-coherence events with the virtually addressed ObjID buffer while preserving its ability to perform ID lookup without physical-address translation on the critical memory-access path.

\section{System and threat model}
As with Picasso and \cherid{}, \reinc{} strengthens temporal safety relative to Cornucopia Reloaded by protecting against both use-after-free (UAF) and use after-reallocation (UAR).
Cornucopia Reloaded prevents UAR by delaying freed memory reuse until dangling capabilities are revoked by a memory sweep, but it does not prevent access to an object while it remains in quarantine.
\reinc{} instead revokes capabilities immediately on free using its ID mechanism, trapping dereference of dangling capabilities. 

\subsection*{Trust model and software topology}
\reinc{} retains the software topology and trust model of Cornucopia Reloaded: the Malloc revocation shim (MRS)~\cite{Cornucopia_reloaded} and allocator remain user-space components.
 
MRS maintains capabilities covering the underlying allocator-managed allocations and returns capabilities tightly bound to application-requested object sizes. The ID metadata lies outside of the application-accessible bounds, while MRS performs ID management using ordinary loads and stores using allocator capabilities. This arrangement does not require new privileged instructions, but still maintains ID integrity in application code. 

Following prior work~\cite{
Cornucopia_reloaded,wangCHERID2026,gulmezPICASSOScalingCHERI2026,
wesleyfilardoCornucopiaTemporalSafety2020}, \reinc{} focuses on user-space heap temporal safety. 
Extending ID support to kernel and stack memory is left for future work.

\section{\reinc{} Design}

\begin{table}[t]
\centering
\caption{Object-ID modes in \reinc{}. }
\label{tab:reinc-idmode}
\small
\begin{tabular}{c|c|c|c}
\hline
\textbf{Mode} &
\textbf{Object size} &
\textbf{ID placement} &
\textbf{IDLOC gran.} \\
\hline
0 & $<63$\,B
  & 128 B region-relative
  & 16\,B \\
1 & 63\,B--4\,KiB
  & Page-relative
  & 1\,B \\
2 & 4--32\,KiB
  & Top-relative
  & 1\,KiB \\
3 & 32--256\,KiB
  & Top-relative
  & 8\,KiB \\
4 & 256\,KiB--2\,MiB
  & Top-relative
  & 64\,KiB \\
5 & 2--16\,MiB
  & Top-relative
  & 512\,KiB \\
6 & 16--128\,MiB
  & Top-relative
  & 4 MiB \\
7 & 128 MiB--1\,GiB
  & Top-relative
  & 32 MiB \\

\hline
\end{tabular}
\end{table}

\reinc{} extends \cherid{} with additional \emph{IDMODEs} to support immediate memory reuse for objects of up to 1 GiB and introduces \emph{allocation reincarnation}, allowing multiple IDs to be dynamically associated with an allocation slot over time.
ID reassignment allows IDs to be quarantined for reclamation while allowing the underlying memory to remain in use.
The \reinc{} \emph{IDLOC} encoding facilitates reuse by placing ID locations in pairs.
These innovations allow continuous reuse of allocation slots in the common case.

\reinc{} further extends \cherid{} by specifying strict coherence for ID buffering without fence instructions.
This is both necessary to maintain safety in multicore systems, and convenient for programmers.

\subsection{\reinc{} architecture extension}
\begin{figure}[t]
        \centering
        \includesvg[width=0.45\textwidth]{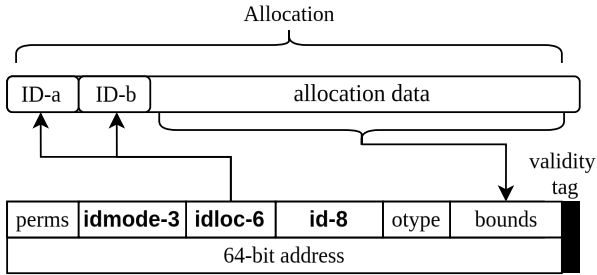}
        \caption{CHERI capability extended with \reinc{} support}
        \label{fig:id_cap_format}
\end{figure}

\reinc{} encodes an 8-bit ID as well as an address of an expected-to-match in-memory ID.
The \reinc{} capability format is shown in Figure~\ref{fig:id_cap_format}.

The \reinc{} architectural extension includes the following:
\begin{itemize}
    \item An 8-bit capability ID field represents the lifetime that a capability is authorized to access, as shown in Figure~\ref{fig:id_cap_format}. 
    \item A 3-bit capability ID mode (IDMODE) field differentiates capabilities using the inline ID storage scheme, the in-page scheme of the original \cherid{}, and the large inline-object scheme~\ref{subsec:large_inline}.
    \item A 6-bit capability ID location (IDLOC) field encodes the memory ID location according to the scheme selected by IDMODE. 
\end{itemize}

ID \emph{0} is reserved for unchecked, ``immortal" caps, such as MRS-maintained underlying capabilities; all user heap capabilities bear non-0 IDs. Updating the capability \emph{ID}, \emph{IDMODE} and \emph{IDLOC} fields is only allowed to be conducted on ID 0 capabilities.

\subsection{Large inline object ID design}
\label{subsec:large_inline}
To extend the object-ID support beyond the 4 KiB boundary of the original \cherid{} design, \reinc{} introduces six additional \emph{IDMODEs} (2-7), along with a new encoding of \emph{IDLOC} that enables the hardware to precisely locate an object's ID metadata in these \emph{IDMODEs}. The original encoding of \cherid{} relies on naturally occurring hardware boundaries, particularly cache-line and page boundaries, to derive the location of an object's ID.

Modes 2-7 introduce a new top-relative ID representation for larger objects that span multiple pages. 
As the supported object size increases, these modes use progressively coarser alignment for ID-location pairs.
Table~\ref{tab:reinc-idmode} summarizes the object-size range and alignment granularity supported by each. 

\reinc{} derives the ID location relative to the capability's upper bound(\emph{top}) under \emph{IDMODEs}(2-5). For these modes, the object's ID location is placed immediately beyond the application-visible capability top, often within the internal fragmentation (unused space) in the underlying allocation slot. This organization follows the same principle as the inline-ID scheme of the original \cherid{} design: ID metadata is co-located with the underlying allocation; it lies outside of the bounds of the application-visible capability and is safe from modification by the application. 

\subsubsection*{Locality-Preserving ID Placement}
This top-relative, co-located ID organization is preferred to maintaining IDs in a separate shadow-memory structure or external metadata table because it preserves the spatial locality between an object and its temporal metadata. When object accesses and their corresponding ID lookups target nearby memory, ID metadata naturally benefits from existing memory optimizations, avoiding additional \textbf{cache and TLB pressure}. For example, \reinc{}'s top-relative organization allows ID lookup for multi-page objects to benefit from superpage mappings that can include both data and ID in a single TLB entry. 

\subsubsection*{Concurrency-friendly ID placement.}
Programmers carefully arrange data structures to avoid contention in concurrent programs.
\reinc{} also naturally respects these optimizations, avoiding a new centralized metadata structure on which ID accesses from multiple cores would contend.

\subsubsection*{Handling Exceptionally Large Objects}
Extending ID-based protection beyond 1 GiB would require additional capability-encoding space for allocations that are rare in measured workloads. 
\reinc{} instead immediately unmaps such allocations upon deallocation, causing subsequent accesses through dangling capabilities to fault.
Unmapping pages at this granularity is particularly efficient as they are likely to be unmappable at the super-page (2MiB) or giga-page (1GiB) granularities.
This hybrid approach allows \reinc{} to dedicate the limited capability encoding space to the object sizes for which efficient ID-based reuse is most beneficial while relying on page-level virtual-memory protection for exceptionally large allocations. 

\subsection{Allocation reincarnation}

\reinc{} allows an allocation slot to be associated with multiple successive IDs, extending the number of temporal generations through which it can be safely reused.
Each 8-bit ID provides up to 255 generations, with \emph{IDLOC} selecting among ID locations that can be associated with an allocation slot.
As in \cherid{}, deallocation advances the selected memory ID, immediately revoking capabilities from the previous generation.
Rather than quarantining memory on ID exhaustion, as in \cherid{},
\reinc{} \emph{reincarnates} the allocation slot.
When an ID reaches 255, IDLOC is updated to select a non-exhausted ID associated with the same allocation slot, providing a fresh sequence of temporal generations.
Capabilities belonging to previous generations remain associated with the old ID and remain revoked while
the underlying memory continues to be reused without being quarantined.

\subsection{ID Quarantine and Reclamation}

Allocation reincarnation decouples \emph{ID exhaustion} from \emph{memory quarantine}. 
Allocation slots can be reincarnated using a new ID location associated with the same slot when the current ID location is exhausted.
Memory quarantine is required only when all IDs available to an allocation slot are exhausted, and no further reincarnation is possible. 

Exhausted IDs, however, do not remain unavailable indefinitely. \reinc{} complements allocation reincarnation with an \emph{ID quarantine and reclamation} mechanism. When an ID reaches its terminal generation, \reinc{} quarantines the exhausted ID.
ID quarantine temporarily removes the exhausted ID location from circulation until
a memory sweep invalidates all outstanding capabilities pointing to quarantined IDs. After the sweep completes, \reinc{} reclaims ID locations by resetting their generation state, making them available for future use. A frequently reused allocation slot can therefore cycle through its current ID while previously exhausted IDs progress through quarantine, sweeping invalidation, and reclamation. This strategy is fundamentally superior to a wider ID field; \reinc{} allows for an effectively infinite temporal generation space without a larger ID. 

ID quarantine substantially reduces quarantine-induced memory pressure by allowing memory to remain in use indefinitely.
This property is particularly important for long-running, allocation-intensive workloads, where quarantine can cause continuous, high memory overhead.

\subsection{ID Placement and Lookup}
\label{sec:id-placement}

As shown in Table~\ref{tab:reinc-idmode}, modes 0 and 1 retain the ID-placement schemes of \cherid{} for small objects. In mode 0, \emph{IDLOC} selects an ID address anywhere in a 64-byte cache line shared with the allocation. 
In mode 1, \emph{IDLOC} selects an ID address from the top 64-bytes of the 4 KiB page that contains the allocation. Because both modes use fixed, address-derived regions, capability bounds narrowing does not require explicit tracking of changes to the capability top.

For larger objects, Modes 2--7 locate IDs relative to the capability upper bound (\emph{top}), using a granularity of 1 KiB, 8 KiB, 64 KiB, 512 KiB, 4 MiB, and 32 MiB, respectively.
\texttt{IDLOC[4:0]} encodes the displacement from the aligned top to the ID metadata. 
The progressively coarser granularity allows the same five-bit displacement to scale to increasingly large objects, supporting allocations of up to 1 GiB. \texttt{IDLOC[5]} independently selects between the two adjacent IDs used for allocation reincarnation.

Algorithm~\ref{alg:idlookup} summarizes ID-address reconstruction and the updates required to preserve the ID location when capability bounds are narrowed. 

\begin{algorithm}[t]
\caption{\reinc{} ID address reconstruction and \emph{IDLOC} preservation. $A$ denotes the access address, $T$ the capability top, $M$ the \emph{IDMODE}, and $L$ the six-bit \emph{IDLOC} field. $\operatorname{AlignDown}(x,g)$ rounds $x$ down to the nearest $g$-byte boundary.}
\label{alg:idlookup}
\begin{algorithmic}[1]

\Function{Granularity}{$M$}
    \State \Return
    $\{1\,\mathrm{KiB},8\,\mathrm{KiB},64\,\mathrm{KiB},
    512\,\mathrm{KiB},4\,\mathrm{MiB},32\,\mathrm{MiB}\}[M-2]$
\EndFunction

\Statex

\Function{IDAddr}{$A,T,M,L$}
    \If{$M = 0$}
        \State $R \gets \operatorname{AlignDown}(A,64\,\mathrm{B})$
        \State \Return $R + 16\,\mathrm{B}\times L[4{:}0]
            - 1 - L[5]$

    \ElsIf{$M = 1$}
        \State $R \gets \operatorname{AlignDown}(A,4\,\mathrm{KiB})$
        \State \Return $R + 4032\,\mathrm{B} + L$

    \Else
        \State $G \gets \Call{Granularity}{M}$
        \State $R \gets \operatorname{AlignDown}(T,G)$
        \State \Return $R + L[4{:}0]\times G - 1 - L[5]$
    \EndIf
\EndFunction

\Statex

\Function{UpdateIDLoc}{$T_{\mathrm{old}},T_{\mathrm{new}},M,L$}
    \State $G \gets \Call{Granularity}{M}$
    \State $R_o \gets
        \operatorname{AlignDown}(T_{\mathrm{old}},G)$
    \State $R_n \gets
        \operatorname{AlignDown}(T_{\mathrm{new}},G)$

    \If{$R_n < R_o$}
        \State $\Delta \gets (R_o-R_n)/G$
        \State $L'[4{:}0] \gets L[4{:}0]+\Delta$
        \State $L'[5] \gets L[5]$
        \State \Return $L'$
    \Else
        \State \Return $L$
    \EndIf
\EndFunction

\end{algorithmic}
\end{algorithm}

\paragraph{Preserving ID location across bounds narrowing}
Top-relative placement must account for CHERI capability derivation:
narrowing capability bounds can reduce the capability top and would
otherwise cause a derived capability to reconstruct a different ID
address. \reinc{} therefore increments \texttt{IDLOC[4:0]} by the number
of mode-specific granularity boundaries crossed as the top decreases,
preserving the original ID location. \texttt{IDLOC[5]} remains unchanged,
keeping reincarnation selection independent of bounds narrowing.

\subsection{Coherence-Aware ID Management}
\label{subsec:coherence management}
\reinc{} must maintain Object-ID (ObjID) coherence across cores to preserve temporal safety. When one core updates an object's ID, other cores may retain stale IDs in their private ObjID buffers; without invalidation, these stale entries could continue to validate revoked capabilities. Subsequent accesses must therefore observe the updated ID through the coherent memory hierarchy.

The most straightforward approach to ensure coherence is to enforce an inclusive policy between the L1 data cache and the ObjId buffer.
If the ObjId buffer only holds IDs that are currently present in the coherent L1 data cache, then we can be certain that no newer value of the ID exists in the system.

Unfortunately, this is challenging because the ObjID buffer is virtually addressed, and cache-coherence events operate on physical addresses. 
To connect physical coherence events to the virtually addressed ObjID buffer without affecting its critical lookup path or substantially modifying the memory subsystem, \reinc{} reuses existing coherence-invalidation and local L1-replacement events.

We explore two approaches. The first maintains a reverse map from physical metadata lines to virtual ObjID locations, enabling selective invalidation of affected entries. In the corner case that conflicting virtual mappings are found to map to the same physical line, indicate aliasing, we conservatively trigger a full-buffer flush.

The second approach combines the structured ID layout with a lightweight Bloom-style physical-address filter. Events that cannot affect cached IDs are ignored, while possible matches conservatively flush the ObjID buffer. The implementations of these mechanisms are described in section~\ref{subsec:coherent}

\section{\reinc{} Implementations}
\subsection{\reinc{} software support}
\reinc{} extends CheriBSD's Malloc Revocation Shim (MRS), which wraps the system allocator to provide heap temporal safety through quarantine and capability revocation. 
MRS intercepts deallocations and quarantines freed objects until the quarantine reaches its reclamation threshold, triggering a sweep that invalidates dangling capabilities before the memory is returned to the allocator. 
\reinc{} modifies this path to immediately revoke freed objects by incrementing their ID, and allowing immediate reuse of the allocation slot.
When the slot is found to be exhausted, MRS performs allocation reincarnation with ID-level quarantine, and finally triggers reclamation when the sweep threshold is reached.

\subsubsection*{ID Quarantine and Reclamation through Sweeping}
\reinc{} extends the MRS (Cornucopia Reloaded) quarantine mechanism to quarantine an exhausted ID independently of its underlying allocation. 
On \textit{free}, when the current ID reaches its terminal value, which is defined as \textit{254}, \reinc{} marks it as exhausted by setting it to \textit{255} and checks whether another ID associated with the allocation remains available. 
If so, only the exhausted ID is inserted into the quarantine, recording its \emph{IDLOC}, while the allocation is immediately returned to the allocator for subsequent reincarnation. 
Each quarantined ID also contributes the size of its associated allocation to a \emph{virtual quarantine size}. 
This accounting allows the exhausted ID to contribute to the existing quarantine threshold and hence to trigger revocation, without withholding the corresponding memory from the allocator. 

Quarantined IDs are reclaimed using \cherid{}'s ID-based invalidation sweep algorithm without relying on the shadow bitmap used for Cornucopia Reloaded. During a memory sweep, the ``revoker" examines capabilities carrying object IDs and invalidates a capability if its associated ID has reached the value \textit{255}. 
Thus, \reinc{} can identify and invalidate capabilities using exhausted ID locations
while the allocation slot itself has already been reincarnated under another ID. 

After the sweep completes, \reinc{} reclaims each quarantined ID by temporarily selecting its recorded \emph{IDLOC} and resetting the corresponding quarantined ID to zero, making the ID available for future reincarnations. If all IDs associated with an allocation slot are exhausted before ID reclamation, the allocation slot itself is quarantined; after a memory sweep, all of its IDs are reset and the memory is again returned to the allocator. 

\subsubsection{Allocator-enforced alignment}
\label{subsubsec:alignment}
To satisfy the alignment requirements of the top-relative \emph{IDMODEs}, \reinc{} requests appropriately aligned allocations through the allocator's aligned allocation interface. This allows the required memory layout to be enforced without modifying the allocator's internal placement mechanisms. Importantly, the required alignment is smaller than the allocator alignment for each size class to ensure that IDs can be placed within internal fragmentation if any is available.
The resulting placement constraint is therefore modest and imposes minimal additional pressure on memory. 

\subsubsection*{Immediate unmapping of super-large allocations}
Besides the allocator modifications required by \cherid{} to support \emph{IDMODE}~1, \reinc{} requires that very large allocations be immediately unmapped upon deallocation. For objects beyond the size range supported by \reinc{}'s ID-based protection, we leverage the operating system's virtual-memory protection to provide use-after-free mitigation. Specifically, page-aligned allocations of at least 1 GiB are returned directly to the operating system upon deallocation using \texttt{pages\_unmap()} to remove their virtual-memory mappings. As these pages can largely be unmapped at super-page (2MiB) or giga-page (1GiB) granularity, this operation should be remarkably efficient. Once \texttt{free()} completes, the virtual address range previously occupied by the allocation is no longer mapped, causing subsequent accesses through dangling capabilities to fault.

\subsection{\reinc{} implementation on CHERI-Toooba}

\subsubsection*{\reinc{} architectural support}
\reinc{} further extends the hardware implementation of \cherid{} with the additional architectural changes that it requires, in particular, updating \emph{TLOC} on upper bound shrinkage and hardware ID lookup for additional \emph{IDMODEs}.

\subsubsection*{Multicore coherent ID buffer}
\label{subsec:coherent}
\reinc{} maintains Object-ID (ObjID) coherence by coupling the core-private ObjID buffer with the existing cache hierarchy of CHERI-Toooba, without modifying the underlying coherence protocol. 
In order to enforce an inclusive policy between the ObjID buffer and the L1 data cache,
ObjID-buffer state must be invalidated both when an L1 line is invalidated by coherence and when it is locally replaced from the L1, since an ID that is not in the L1 data cache may be modified in the system without generating an invalidation observed by this core.

We implement the two coherence mechanisms described in Section~\ref{subsec:coherence management} using the existing L1 coherence-invalidation and local replacement events. Each event provides the physical address of the affected cache line, which is used by either the filter or reverse map to determine whether the ObjID-buffer state must be invalidated.

The \emph{filter-based design} exploits the software-enforced metadata layout described in~\ref{subsubsec:alignment}. For metadata organized at a 1 KiB granularity, only the final 64-byte line of each 1 KiB region can contain IDs,
\begin{equation}
    F_{1\mathrm{K}} =
    \left(\texttt{lineAddr}[3:0] = 4'hF\right),
\end{equation}
while for layouts using a 4 KiB or larger granularity, the relevant line is the final line of a 4 KiB page,
\begin{equation}
    F_{4\mathrm{K}} =
    \left(\texttt{lineAddr}[5:0] = 6'h3F\right).
\end{equation}
We combine this static layout filter with a small Bloom-style physical-address filter, recording cache lines may have corresponding entries in the ObjID buffer. Lines that pass the layout check are queried against the Bloom filter. A miss requires no action, while a hit conservatively flushes the ObjID buffer. False positives are safe and result only in unnecessary flushes. The static filter reduces the number of Bloom filter queries and opportunities for false-positive flushes, while the Bloom filter detects evicted cache lines that might currently be held in the ObjID buffer.
Together, these mechanisms avoid flushes for most unrelated L1 events with little additional hardware, as shown in Table~\ref{tab:fpga-area}.

The \emph{reverse-map design} maintains a small mapping from physical metadata cache lines to their corresponding virtual ObjID locations. On either a coherence invalidation or a local L1 replacement, the physical line is looked up in the reverse map. Since each 64-byte metadata line corresponds to four 16-byte ObjID-buffer entries, the recovered virtual address is aligned to 64 bytes, and the entries at offsets
$\{0,16,32,48\}$ from $V_{\mathrm{line}}$ are invalidated.

ObjID buffer entries are 16 bytes, and a 64-byte data cache line could be spread across four ObjID buffer entries.
If so, these entries are invalidated sequentially, allowing only one virtual address to be stored per reverse-map entry.
If the same physical metadata line is observed through different virtual lines, \reinc{} detects the alias and conservatively clears both the ObjID buffer and reverse-map state.

The two designs trade hardware cost for invalidation precision. The Bloom-filter design requires less hardware but may conservatively flush the entire buffer on a false positive, whereas the reverse map requires additional memory resources but normally invalidates only the affected entries. Both reuse existing coherence-invalidation and local L1 replacement events, require no new coherence messages or protocol states, and leave the timing-critical virtually addressed ObjID lookup path unchanged.

Local L1 replacement is handled conservatively because eviction does not necessarily imply that the underlying ObjIDs have been modified; a line may simply be displaced due to cache pressure. Since the mechanisms cannot determine whether a replacement will be followed by a modification that makes cached ObjIDs stale, they invalidate the corresponding entries to preserve correctness. This conservation may introduce additional ObjID-buffer misses and ID accesses, but it enables coherence through simple, core-local changes without invasive modifications to the memory hierarchy. Future work could eliminate unnecessary invalidation through tighter integration with the existing cache-coherence protocol.

\section{Evaluation}

We evaluate \reinc{} using an FPGA implementation based on
CHERI-Toooba~\cite{ruggSuiteProcessorsExplore2024}. \reinc{} is implemented as a hardware--software co-design spanning the CHERI software stack, including the LLVM compiler toolchain, CheriBSD, and the user-space allocator, together with QEMU and the CHERI-Toooba hardware implementation.

For performance and hardware evaluation, we deploy the modified CHERI-Toooba system on a VCU118 FPGA. 
The system is configured with two cores, each with an 8-way 32\,KB L1 data cache, and a shared 16-way 1\,MB last-level cache connected to the CHERI tag controller and DRAM. We synthesize the design using Vivado 2019 at a clock frequency of 25\,MHz.

We evaluate \reinc{} along four dimensions:

\begin{itemize}
    \item \textbf{Hardware area overhead.}
    We evaluate the hardware area overhead introduced by \reinc{} by reporting the FPGA resource utilization of the \reinc{} extensions relative to the baseline CHERI-Toooba.
    \item \textbf{Temporal safety.}
    We evaluate the temporal-safety guarantees provided by \reinc{} using version 1.3 of the U.S. NIST SARD Juliet Test Suite~\cite{Juliet} and the MSET framework~\cite{vintila2025mset}.

    \item \textbf{Performance.}
    We use SPEC CPU2006 INT, PARSEC, and SQLite workloads~\cite{henningSPECCPU2006Benchmark2006, bienia2008parsec, sqlite}, with execution time overhead as the primary performance metric. 
    
    \item \textbf{Quarantine and sweep behavior.}
    We also collect \reinc{}'s quarantine memory overhead and the number of memory sweeps to compare with \cherid{} as well as Cornucopia Reloaded. 

\end{itemize}

\subsection{Hardware area}

\begin{table}[t]
\centering
\caption{FPGA resource utilization reported by Vivado 2019.}
\label{tab:fpga-area}
\small
\setlength{\tabcolsep}{3pt}
\begin{tabular}{lrrrr}
\toprule
Resource & Baseline & \reinc{} & Reverse & Filter \\
\midrule
LUT Logic
& 261552
& \shortstack{275345\\(+5.27\%)}
& \shortstack{279448\\(+6.84\%)}
& \shortstack{277642\\(+6.15\%)} \\

Registers
& 129099
& \shortstack{140749\\(+9.02\%)}
& \shortstack{146217\\(+13.26\%)}
& \shortstack{144376\\(+11.83\%)} \\

LUT Memory
& 5343
& \shortstack{5823\\(+8.98\%)}
& \shortstack{6575\\(+23.06\%)}
& \shortstack{5842\\(+9.34\%)} \\
\bottomrule
\end{tabular}
\end{table}

As shown in Table~\ref{tab:fpga-area}, both coherence designs introduce limited overall FPGA resource overhead, although the reverse map increases LUT-memory use by 23.06\% relative to the baseline in exchange for selective invalidation. Both designs avoid adding logic to the timing-critical ObjID lookup path.

\subsection{Security}
We evaluate \reinc{} using the NIST SARD Juliet Test Suite~\cite{Juliet} and MSET~\cite{vintila2025mset}. We execute all 2422 applicable CWE-415 (Double Free) and CWE-416 (Use After Free) Juliet tests on our FPGA implementation and use MSET-generated tests to exercise additional temporal-safety patterns.

\subsubsection*{Use-After-Free}
Both our FPGA-based and QEMU-based \reinc{} prototype implementations have successfully executed all provided CWE-416 good test cases and detected and trapped all 416 vulnerabilities in the ``bad'' cases. 
These results demonstrate that the architectural mechanism introduced by \reinc{} enables CHERI to provide true use-after-free mitigation, extending beyond Cornucopia Reloaded's use-after-reallocation mitigation~\cite{Cornucopia_reloaded} and is in the same class as CHERIoT~\cite{CHERIoT}, PoisonCap~\cite{wangPoisonCapEfficientHierarchical2026}, Picasso~\cite{gulmezPICASSOScalingCHERI2026}, and CHERI-D~\cite{wangCHERID2026}.

We also confirm correctness using the 12 generated heap use-after-free cases in the \textbf{MSET} benchmark suite~\cite{vintila2025mset}. 
Of the eight executable cases, four exercise use after reallocation and are rejected by existing capability revocation, while four access memory between deallocation and reuse and are rejected by \reinc{}'s ID validation. 

\subsubsection*{Multicore ID-coherence microbenchmark}
We construct a two-thread ID-coherence test to validate \reinc{}'s coherence  mechanism. This test requires that a remote ID update must be propagated across cores for temporal-safety enforcement to remain correct. Both cores first access the same object, causing its memory ID to be cached in their respective private ID buffers. 
Core 1 retains a capability carrying the current ID, while Core 0 subsequently updates the object's memory ID, emulating the ID transition that occurs during deallocation or reincarnation. Core 1, therefore, holds both a stale capability and, potentially, a stale copy of the object's memory ID in its private ID buffer.

This scenario exposes a limitation of the previous \cherid{} and Picasso hardware in a multicore setting~\cite{gulmezPICASSOScalingCHERI2026, wangCHERID2026}. An ID update performed by core 0 does not invalidate the corresponding state cached in core 1's private ID buffer. Consequently, when core 1 subsequently dereferences its stale capability, the capability's old ID may still match the stale ID retained in its local ID buffer. The access can therefore pass the ID check even though the object's memory ID may have already been changed by another core. In other words, updating an object's ID on one core may not guarantee that stale capabilities to that object will be immediately rejected on another core. 

In \reinc{} hardware, when core 0 modifies an object ID, the write targets an ID-storage region. The resulting coherence invalidation observed by Core 1 triggers its ObjID-buffer coherence mechanism. With the reverse-map design, the corresponding ObjID-buffer entries are selectively invalidated; with the Bloom-filter design, a filter match conservatively flushes the entire ObjID buffer. In either case, the stale ID is removed, ensuring subsequent accesses load an up-to-date ID and trap if they are from stale capabilities.

Both \reinc{} coherence designs consistently detect stale-capabilities, even with remote ID updates, verifying that
\reinc{} successfully enforces the cross-core temporal-safety.

\subsubsection*{Double Free}
We also evaluated our \reinc{} prototype implementations to preserve the double free security properties that \cherid{} preserved, and \reinc{} has successfully passed all 1636 tests in the CWE-415: Double Free class of the Juliet Test Suite~\cite{Juliet}.

\subsection{Performance}
\begin{figure}[htbp]
    \centering
    \includesvg[width=\columnwidth]{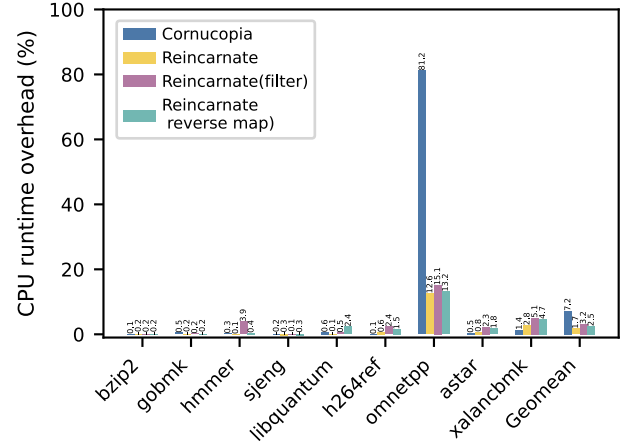}
    \caption{SPEC CPU2006 INT runtime overhead of Cornucopia Reloaded and \reinc{} relative to baseline CHERI without revocation.}
    \label{fig:cpu_cycle}
\end{figure}

\begin{figure*}[t]
    \centering
    \includegraphics[width=\textwidth]{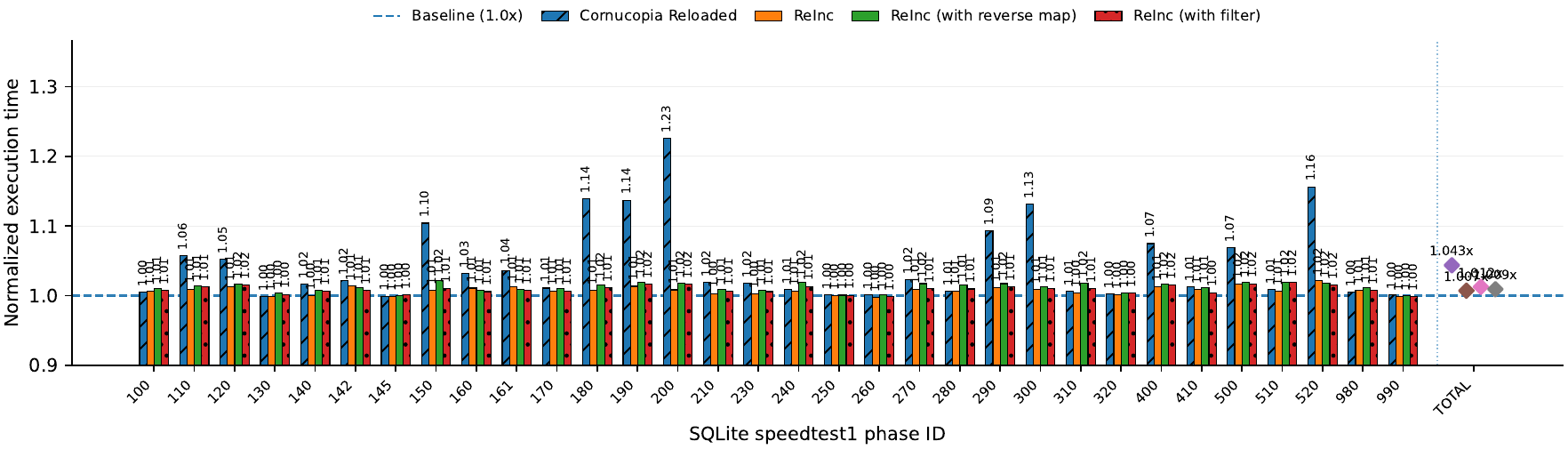}
    \caption{Normalized performance overhead of \reinc{}(0.7\%, 2.1\% maximum), \reinc{} with filter (0.9\%, 1.9\% maximum), \reinc{} with reverse map(1.2\% average, 2.1\% maximum), and Cornucopia Reloaded (4.3\% average, 22.6\% maximum) across the operational phases of SQLite \texttt{speedtest1}.}
    \label{fig:sqlite_perf}
\end{figure*}

\begin{figure*}[t]
    \centering
    \begin{subfigure}[b]{0.47\textwidth}
        \centering
        \includegraphics[width=\linewidth]{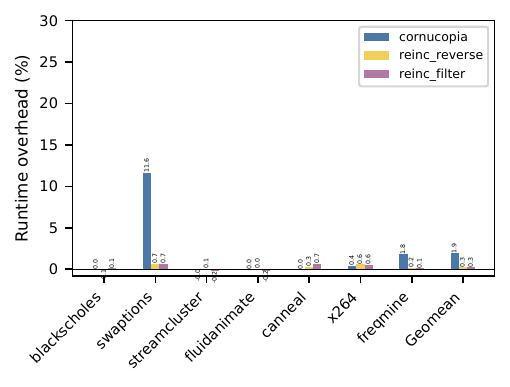}
        \caption{Single-thread runtime overhead}
        \label{fig:parsec-overhead-t1}
    \end{subfigure}
    \hfill
    \begin{subfigure}[b]{0.47\textwidth}
        \centering
        \includegraphics[width=\linewidth]{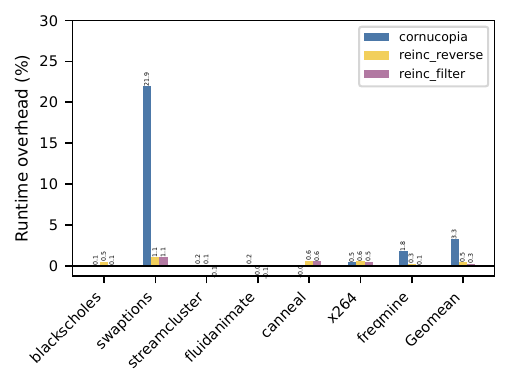}
        \caption{Two-thread runtime overhead}
        \label{fig:parsec-overhead-t2}
    \end{subfigure}
    \caption{PARSEC runtime overhead of Cornucopia Reloaded and \reinc{} relative to the baseline CHERI without revocation.}
    \label{fig:parsec-overhead}
\end{figure*}

We evaluate three \reinc{} configurations: without multicore coherence, with the reverse map, and with the lightweight filter.

\subsubsection*{SPEC CPU2006 INT}
To evaluate the performance and memory-utilization improvements of \reinc{} over prior work, we use the CHERI-supported subset of the SPEC CPU2006 INT benchmarks~\cite{henningSPECCPU2006Benchmark2006}.
Consistent with prior performance evaluations on CHERI-Toooba~\cite{ruggEfficientSpatialTemporal, gulmezPICASSOScalingCHERI2026, wangPoisonCapEfficientHierarchical2026}, we primarily use the \emph{train} input configuration. 

\reinc{} substantially reduces the performance overhead of Cornucopia Reloaded, particularly for allocation-intensive workloads such as \textit{omnetpp}, and incurs an average overhead of 1.7\%. 
These improvements stem primarily from immediate memory reuse and reduced sweep frequency, as reflected in Table~\ref{tab:revocation-quarantine-count}. 
For the majority of the benchmarks, \reinc{} incurs 0 revocation and quarantine events. 
The filter-based and reverse-map coherence mechanisms also remain efficient, incurring average overheads of 3.2\% and 2.5\%, respectively. \reinc{} has also substantially reduced the DRAM traffic overhead, incurring only 6\%, whereas Cornucopia Reloaded incurs 48.2\%, as shown in Figure~\ref{fig:llc_miss} in Appendix~\ref{appendix:dram_traffic}.

\subsubsection*{SQLite}

We evaluate SQLite's \texttt{speedtest1}, which exercises allocations across object sizes beyond \cherid{}'s 4 KiB limit. As shown in Figure~\ref{fig:sqlite_perf}, \reinc{} incurs only 0.7\% average runtime overhead, compared with 4.3\% for Cornucopia Reloaded, while reducing the maximum runtime overhead from 22.6\% to 2.1\%. The reverse-map and filter configurations similarly incur average overheads of 1.2\% and 0.9\%, respectively.

This improvement corresponds directly to reduced quarantine and revocation pressures. 
As shown in Table~\ref{tab:revocation-quarantine-count}, Cornucopia Reloaded performs 267 memory sweeps during execution, whereas \reinc{} requires 1 (0 without ID-sweep reclamation). \reinc{} also triggers 90 memory-quarantine events, compared with 221K for Cornucopia Reloaded and 146K for \cherid{}.

\subsubsection*{PARSEC}
We evaluate the overhead of \reinc{} on multithreaded workloads using seven applications from PARSEC~\cite{bienia2008parsec}: \textit{blackscholes, swaptions, streamcluster, fluidanimate, canneal, x264}, and \textit{freqmine}. PARSEC complements the single-threaded SPEC CPU2006 INT evaluation by exercising shared-memory workloads in which \reinc{}'s multicore ID-coherence mechanisms are active. 
Our VCU118 FPGA platform supports at most two CHERI-Toooba cores, each with a single hardware thread; therefore, we evaluate both one- and two-thread executions.

Figures~\ref{fig:parsec-overhead-t1} and \ref{fig:parsec-overhead-t2} report the one- and two-thread results, respectively. \reinc{} maintains low overhead when moving from single-threaded to two-threaded execution, including when either multicore coherence mechanism is enabled. The filter-based design performs comparably to the more precise reverse map design despite conservatively flushing the ObjID buffer on possible matches.

As shown in Table~\ref{tab:revocation-quarantine-count}, Cornucopia Reloaded incurs low overhead for most PARSEC workloads, but \textit{swaptions}' high allocation and reuse rate triggers frequent revocation sweeps, resulting in substantially higher overhead.
\reinc{} instead reduces revocation pressure through allocation reincarnation and ID-level quarantine, retaining low overhead for this workload.

Our FPGA resource budget limits this evaluation to two cores; while these results nevertheless demonstrate low multicore coherence overhead, we were not able to evaluate many-core scalability.
However, as IDs are distributed with data and propagate using traditional cache coherence, we do not expect performance scaling to deviate greatly from the baseline.

\subsection{Reducing memory-quarantine overhead}

\begin{table}[t]
\centering
\caption{
Allocation, revocation, and memory-quarantine behavior.
The benchmark column gives allocation count in parentheses.
Entries are \emph{S/Q}, where S and Q denote revocation sweeps
and memory-quarantine events, respectively.
For \reinc{}, a bracketed S value gives the sweep count when
ID-sweep reclamation is disabled.
For SPEC CPU2006, ``tr'' and ``ref'' denote the train and
reference input sets, respectively.
}
\label{tab:revocation-quarantine-count}

\small
\setlength{\tabcolsep}{2.5pt}
\renewcommand{\arraystretch}{1.03}

\begin{tabular}{@{}l r r r@{}}
\toprule
\textbf{Benchmark (Alloc.)}
& \textbf{Corn.}
& \textbf{\cherid{}}
& \textbf{\reinc{}} \\
\midrule

\multicolumn{4}{@{}l}{\textit{SPEC CPU2006}} \\

bzip2-tr (30)
& 2/24
& 2/24
& 0/0 \\

bzip2-ref (30)
& 0/24
& 0/24
& 0/0 \\

gobmk-tr (51K)
& 11/51K
& 6/28K
& 0/0 \\

gobmk-ref (133K)
& 23/133K
& 2/82K
& 0/0 \\

hmmer-tr (170K)
& 16/170K
& 1/693
& 0/134 \\

hmmer-ref (1.0M)
& 124/1.0M
& 2/3.9K
& 0/782 \\

sjeng-tr (6)
& 0/1
& 0/1
& 0/0 \\

sjeng-ref (6)
& 5/1
& 0/1
& 0/0 \\

libquantum-tr (109)
& 1/107
& 1/58
& 0/0 \\

libquantum-ref (150)
& 11/149
& 11/57
& 0/0 \\

h264ref-tr (38K)
& 12/38K
& 12/5.1K
& 0/0 \\

h264ref-ref (38K)
& 12/38K
& 12/5.1K
& 0/0 \\

omnetpp-tr (130M)
& 2792/130M
& 12/151K
& 15[2]/99K \\

omnetpp-ref (267M)
& 608/267M
& 6/304K
& 0/200K \\

astar-tr (347K)
& 38/347K
& 9/2.7K
& 0/47 \\

astar-ref (1.1M)
& 187/1.1M
& 27/13K
& 0/94 \\

xalancbmk-tr (1.1M)
& 2/1.1M
& 0/34K
& 0/1K \\

xalancbmk-ref (135M)
& 291/135M
& 9/2.5M
& 0/178K \\

\midrule
\multicolumn{4}{@{}l}{\textit{SQLite}} \\

speedtest1 (221K)
& 267/221K
& 242/146K
& 1[0]/90 \\

\midrule
\multicolumn{4}{@{}l}{\textit{PARSEC}} \\

blackscholes (14)
& 1/6
& 1/3
& 0/0 \\

swaptions (19.7M)
& 9516/19.7M
& 7384/6.6M
& 7/27K \\

streamcluster (1.8K)
& 0/1.8K
& 0/8
& 0/2 \\

fluidanimate (15K)
& 1/15K
& 1/10
& 0/0 \\

canneal (872)
& 0/866
& 0/3
& 0/0 \\

x264 (270K)
& 11/270K
& 0/375
& 0/353 \\

freqmine (611)
& 3/592
& 2/562
& 0/0 \\

\bottomrule
\end{tabular}
\end{table}

\begin{figure*}[t]
    \centering

    \begin{subfigure}[t]{0.49\textwidth}
        \centering
        \includesvg[width=\linewidth]{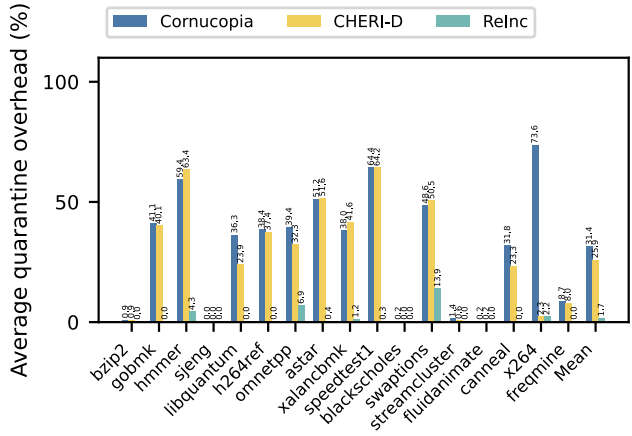}
        \caption{Time-weighted average memory-quarantine occupancy.}
        \label{fig:quarantine-avg}
    \end{subfigure}
    \hfill
    \begin{subfigure}[t]{0.49\textwidth}
        \centering
        \includesvg[width=\linewidth]{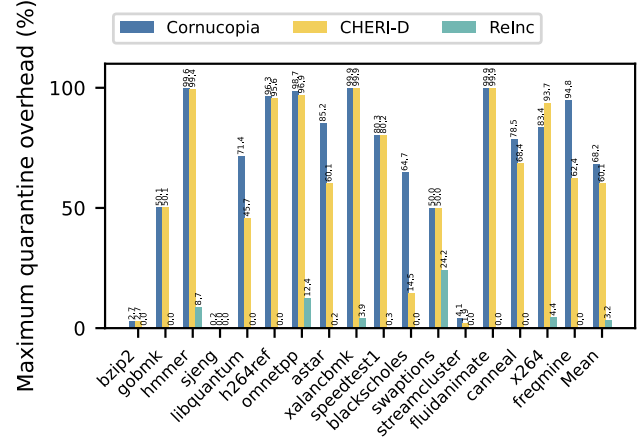}
        \caption{Maximum memory-quarantine occupancy.}
        \label{fig:quarantine-max}
    \end{subfigure}

    \caption{
    Quarantine memory overhead across the evaluated workloads.
    Average quarantine overhead is the time-weighted average of $Q(t)/A(t)$ over execution, where $Q(t)$ is memory-quarantine occupancy and $A(t)$ is allocated memory at time $t$.
    Maximum quarantine overhead (Max-Q) is
    $Q(t_q)/A(t_q)$ at the point $t_q$ of maximum memory-quarantine occupancy. 
    Values are percentages. SPEC CPU2006 results use the reference inputs.
    }
    \label{fig:quarantine-overhead}
\end{figure*}
In addition to performance improvements, \reinc{} substantially reduced the amount of memory held in quarantine through \emph{allocation reincarnation}. 
As shown in Table~\ref{tab:revocation-quarantine-count}, \reinc{} incurs significantly fewer memory-quarantine events than Cornucopia Reloaded and \cherid{}, with no memory-quarantine events for the majority of the evaluated benchmarks. 

Quarantine-event count alone does not capture the amount of memory withheld from reuse. We therefore measure both peak memory-quarantine occupancy and its time-weighted average over execution. 
Figure~\ref{fig:quarantine-avg} and Figure~\ref{fig:quarantine-max} show that \reinc{} substantially reduces peak quarantine memory overhead.
ID quarantine and reclamation mechanisms not only reduce how often memory enters quarantine, but also substantially reduce the peak memory in quarantine, as a large proportion of the quarantine budget is occupied by allocation slots that were successfully reincarnated and are not consuming memory.

\section{Related work}
\subsubsection*{Use-after-reallocation mitigation}
CHERIvoke introduced sweeping capability revocation for CHERI, with Cornucopia, Cornucopia Reloaded, and CHERIoT subsequently developing related quarantine-and-revocation mechanisms~\cite{CHERIvoke, wesleyfilardoCornucopiaTemporalSafety2020, Cornucopia_reloaded}.
These approaches generally prevent the immediate reuse of freed memory by quarantining it until dangling capabilities are identified and revoked.
\reinc{} improves upon Reloaded by enforcing strict use-after-free detection and allowing immediate memory reuse.

Software-based approaches have also explored temporal memory safety. 
DangNull nullifies dangling pointers during object deallocation~\cite{dangnull}, while CETS maintains metadata to validate pointer dereferences at runtime~\cite{cets}. 
MarkUs and Minesweeper instead provide temporal safety through software-based memory-management and reclamation mechanisms~\cite{ainsworthMarkUsDropinUseafterfree2020,erdosMineSweeperCleanSweep2022}. Managed-memory systems avoid dangling pointers through automatic lifetime management; garbage collectors like Oilpan~\cite{oilpan} in Chromium's Blink engine use this technique.

\subsubsection*{Memory tagging}
Arm's Memory Tagging Extension (MTE) associates tags on pointers with out-of-band tags on each memory word. 
With only four tag bits, frequent memory reuse inevitably causes tag aliasing, providing only probabilistic temporal safety. 
Combining MTE with Cornucopia Reloaded can provide deterministic temporal safety~\cite{DiscussionMSRsCHERI+MTE}, but limits reuse to 15 generations before quarantine. 
Moreover, because MTE tags memory at a fixed granularity, changing a large object's tag requires updating tags throughout the object, with cost increasing with object size.

In contrast, \reinc{} uses per-allocation IDs so that an entire object is invalidated with a single write; and uses reincarnation so that exhausted IDs can be reclaimed while memory remains in use.

\subsubsection*{Centralized hardware metadata tables}
CHERI has generally followed a decentralized design philosophy, avoiding centralized metadata structures and additional indirection.
CHERIoT and Picasso instead maintain temporal metadata in centralized hardware tables. 
While practical for CHERIoT's single-core embedded setting~\cite{CHERIoT}, this organization is harder to scale to application-class multicore systems with large, potentially NUMA, memories. 
Separating metadata from object data also reduces locality and may require additional address translation during validation, increasing TLB and metadata-cache pressure.

Picasso reduces memory-sweep frequency using a 21-bit capability color, but consumes substantial capability-encoding space; 
using fewer bits would increase color-reuse and revocation
pressure~\cite{gulmezPICASSOScalingCHERI2026}.  Together with its centralized metadata, this complicates scaling to allocation-intensive multicore systems.
\reinc{} 
requires only 17 capability bits to achieve full scalability with decentralized metadata storage.

\subsubsection*{Memory poisoning} 
PoisonCap extends Cornucopia Reloaded to provide strict, hierarchical use-after-free mitigation on CHERI~\cite{wangPoisonCapEfficientHierarchical2026}. It introduces \emph{Poison capabilities}, which are written into freed memory and cause subsequent accesses to trap. 
Unlike \reinc{}, PoisonCap does not enable immediate reuse of freed memory; instead, it reduces the microarchitectural cost of memory quarantine by improving its cache efficiency.
While PoisonCap does not substantially reduce memory-sweep frequency as \reinc{} does, it provides a broader security model, supporting strict use-after-free protection across multiple allocation layers and also provides initialization safety. 
The two approaches address complementary aspects of CHERI temporal safety: PoisonCap strengthens protection and reduces the microarchitectural cost of quarantine, whereas \reinc{} reduces the need for memory quarantine and revocation by enabling immediate memory reuse, ID quarantine with reincarnation, and ID reclamation.

\section{Future work}

\subsubsection*{Hierarchical temporal memory safety}
\reinc{} currently protects heap allocations at the libc allocator layer.
As highlighted by PoisonCap, however, use-after-free vulnerabilities can also arise at other allocation layers, including kernel allocators and application-specific nested allocators~\cite{wangPoisonCapEfficientHierarchical2026}. 
Extending \reinc{} to provide scalable temporal protection across multiple allocation layers, while preserving its decentralized ID management and efficient allocation reincarnation, is an important direction for future work.

\subsubsection*{\reinc{}-integrated memory allocator} Our current \reinc{} software support is implemented in the MRS wrapper rather than directly in the underlying memory allocator. This introduces additional overhead because MRS must recover the underlying allocation and associated metadata to locate and update object IDs during allocation and deallocation. 
A native \reinc{}-aware allocator could instead perform ID selection, update, quarantine, and reclamation directly using allocator metadata already available on these critical paths, eliminating much of this wrapper overhead.

\section{Conclusion}
\reinc{} demonstrates that strong temporal memory safety can be achieved with substantially lower memory-sweep frequency and quarantine memory overhead. 
Its key mechanism, \emph{allocation reincarnation}, allows an allocation slot whose current ID is exhausted to be immediately reincarnated with a new ID location, rather being placed in quarantine. 
\reinc{} also supports efficient cross-core synchronization of the IDs which extends temporal safety support on multi-core systems. 
\reinc{} thus decouples temporal-identity reuse from memory reuse, providing a scalable approach to temporal safety for application and server class CHERI systems while preserving CHERI's decentralized architecture.
These advancements demonstrate that temporal safety can be enforced on CHERI systems not only low cycle overhead, but very low memory overhead as well.

\clearpage

{
\raggedright
\bibliographystyle{ACM-Reference-Format}
\balance
\bibliography{software}
}

\clearpage
\appendix

\section{SPEC DRAM traffic overhead}
\label{appendix:dram_traffic}
\begin{figure}[htbp]
    \centering
    \includesvg[width=\columnwidth]{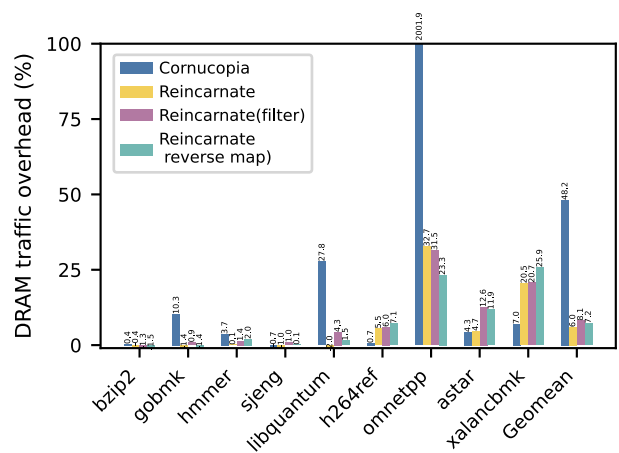}
    \caption{SPEC CPU2006 INT DRAM traffic overhead of Cornucopia Reloaded and \reinc{} relative to baseline CHERI without revocation.}
    \label{fig:llc_miss}
\end{figure}

\end{document}